\documentclass[12pt]{revtex4}

\usepackage{dcolumn}
\usepackage{bm}
\usepackage{color}

\usepackage{amssymb} 
\usepackage{graphicx}
\usepackage{amsmath}

\newcommand{\be}{\begin{equation}}
\newcommand{\ee}{\end{equation}}
\newcommand{\ud}{\mathrm{d}}

\begin{document}

\pagestyle{plain}

\title{
\vskip -120pt
{\begin{normalsize}
\mbox{} \hfill DAMTP-2007-16\\
\vskip  30pt
\end{normalsize}}
{\bf\Large
Observing cosmic string loops \\ with gravitational lensing surveys
}
}

\author{Katherine J. Mack$^{1,2,3}$}
\email{mack@astro.princeton.edu}
\affiliation{$^1$Department of Astrophysical Sciences, Princeton University}

\author{Daniel H. Wesley$^2$}
\email{D.H.Wesley@damtp.cam.ac.uk}
\affiliation{$^2$Department of Applied Mathematics and Theoretical Physics, Cambridge University}

\author{Lindsay J. King$^3$}
\email{ljk@ast.cam.ac.uk}
\affiliation{$^3$Institute of Astronomy, Cambridge University}

\begin{abstract}
\noindent 
We show that the existence of cosmic strings can be strongly constrained
by the next generation of gravitational lensing surveys at radio
frequencies.  We focus on cosmic string loops, which simulations
suggest would be far more numerous than long (horizon-sized) strings.
Using simple models of the loop population and minimal assumptions about
the lensing cross section per loop, we estimate the optical depth to
lensing and show that extant radio surveys such as CLASS have already
ruled out a portion of the cosmic string model parameter space.  Future
radio interferometers, such as LOFAR and especially SKA, may constrain
$G\mu/c^2 < 10^{-9}$ in some regions of parameter space, outperforming
current constraints from pulsar timing and the CMB by up to two orders of
magnitude.
This method relies on direct detections of cosmic strings, and so is
less sensitive to the theoretical uncertainties in string network
evolution that weaken other constraints.

\end{abstract}

\maketitle

\section{Introduction}

Cosmic strings are linear, relativistic objects whose existence is 
predicted by a number of extensions to the Standard Model of particle 
physics
\cite{Kibble1976,HindmarshKibble1995,VilenkinShellard2000}.
When first proposed, they were intensely studied as a possible 
mechanism for cosmological structure formation.  This
scenario is now 
known to be inconsistent with a variety of cosmological constraints \cite{CaldwellAllen1992,Caldwelletal1996,LandriauShellard2004,Pogosianetal2004,Pogosianetal2006,JeongSmoot2005,Pogosianetal2003,PenSeljakTurok1997,Allenetal1997,Albrechtetal1999,Spergeletal2006},
but recently cosmic strings have attracted renewed interest 
thanks to the postulation of new types and new production mechanisms.  
The current focus has shifted from their role in structure formation to 
the prospects for detecting them if they are present
\cite{Pogosianetal2004,Pogosianetal2006,Wymanetal2006,Sazhinetal2006,ShlaerWyman2005,SeljakSlosar2006}. 

Here we describe a method to detect cosmic string loops using strong 
gravitational lensing of compact radio sources.  We propose searching
for loops because simulations suggest there could be $>10^5$ per
horizon volume in contrast to only $\sim 40$ long (horizon-sized)
strings \cite{HindmarshKibble1995,VilenkinShellard2000}.
Compact radio sources (hereafter CRSs) are an ideal source population
since their 
point-like nature on angular scales $< 100$ mas makes lensed images
easy to identify. Furthermore, numerous CRSs have been observed 
by radio surveys such as the Cosmic Lens All-Sky Survey (CLASS) and are 
expected to be observed by proposed surveys such as the Low Frequency
Array (LOFAR) and the Square 
Kilometer Array (SKA).  In this paper, we estimate the 
expected number of loop lensing events for these three surveys as a 
function of cosmic string model parameters, and find that for 
significant regions of parameter space our method can outperform existing 
constraints.

An important feature of our method is that it would constitute a
{\em direct detection} rather than an inference based on assumed or
simulated properties of a cosmic string network.  By contrast, the
pulsar timing 
constraint (the most stringent constraint to date) requires knowledge
of the rate of emission of gravitational waves by oscillating strings
and of properties of the loop population.  As we will discuss below,
the signatures of lensing by a string loop are distinctively different
from those of lensing by a galaxy (which will generally have a similar
image separation in the arcsecond regime).
Follow-up observations should be able to
quickly determine if putative lensing events are due to cosmic strings
or to more conventional astronomical objects.  In past gravitational
lensing surveys, all lensing events have been associated with 
conventional lensing objects such as galaxies or clusters after 
follow-up observations.  We rely on the reasonable expectation that
sufficient follow-up is always performed; a constraint will be obtained
if all lensing events are accounted for by standard lenses and are
therefore not cosmic string lensing candidates.  If after deep imaging
a conventional lensing object was
not observed, the lens would be classified as ``dark" and modelling of the gravitational potential 
would be required to rule it out as a cosmic string.  As of yet, no confirmed dark lenses have been found.

Either a detection of, or a constraint on, cosmic strings would be
important for understanding physics beyond the Standard Model. Cosmic
strings can be accommodated in many superstring models, and indeed are
predicted to be present in brane inflation scenarios
\cite{DvaliTye1999,Jonesetal2002,SaswatSarangiTye2002,JonesStoicaTye2003,Witten1985,Copelandetal2004,Polchinski2005}.
The wide range of string tensions and the novel
features of cosmic strings predicted by these models present a
challenge to observation.  By tightening constraints on cosmic
strings, we are indirectly testing these models and learning about
physics at high energies and early times in the history of the
universe. 

This paper is organized as follows.  In Section \ref{s:Methodology} we 
describe our assumptions about the properties of the loop lenses,
discuss our 
population models, and describe the telescopes and surveys proposed
for the observation.
We present the results of the calculation in Section 
\ref{s:Results}, giving the expected number of lensing events as a 
function of parameters of the string model and comparing the constraint 
obtained to those presently available.  
We conclude in Section 
\ref{s:Discussion} and discuss areas in which future work is needed.

\section{Methodology}\label{s:Methodology}

We now describe our calculation of the expected number of observable 
string loop lensing events.  We do this by combining estimates of the 
area of sky that can be strongly lensed by loops,  the resolution and dynamic 
range of the instrument employed, and the source population abundance, which is
a function of instrument and survey parameters as well as of the intrinsic 
abundance.

To estimate of the fraction of the sky that is strongly lensed by the 
loop population we must confront the significant theoretical 
uncertainties in both (i) the lensing characteristics of individual 
loops, and (ii) the distribution of lengths in the loop population as a 
function of redshift. To deal with (i) we take a statistical approach: 
we assume that each loop can form multiple images
of all sources in a patch of sky of angular area 
$\pi\theta_E^2$, with image separations of at least $\theta_E$,
where $\theta_E$ is the Einstein radius of a 
Schwarzschild lens with the same mass as the loop.

Realistic loops 
will be of highly irregular shape and can produce complicated lensing 
patterns, which have been considered in detail elsewhere
\cite{HutererVachaspati2003,UzanBernardeau2000,deLaixVachaspati1996,HoganNarayan1984}.  
However, we expect that the lensing cross section, when averaged over a 
population of loops of fixed mass, will be close to the characteristic 
area set by the Schwarzschild lens.  We justify our approximation in 
Section \ref{ss:LoopsAsLenses} by comparing it to the exact results in 
cases where loop lensing can be studied analytically. To model the loop 
population as required by (ii), we use two simple models, which we 
discuss in Section \ref{ss:LoopPopModels}, with the string tension 
$\mu$ a free parameter in both.  The first is a powerlaw distribution 
of loop lengths, defined by a spectral index $\gamma$, the loop density 
parameter $\Omega_{loop}$, and a cutoff length $L_*$.  The second model 
is derived from the one-scale model, a semi-analytic model of string 
networks widely discussed in the literature.  This model has a number 
of parameters, but in addition to $\mu$ we consider only $\alpha$ (which sets the size of newly created loops) to be free.

We then turn to the source population and instruments.  In Section 
\ref{ss:CRS} we argue that compact radio sources make ideal source 
candidates for investigating lensing by loops, and we discuss some of 
their features.
In Section \ref{ss:RadioSurveys}, we describe both extant surveys (CLASS) 
and proposed instruments (LOFAR and SKA) that will observe these objects.
Our discussion focuses on the parameters of these surveys that are
most relevant to a prediction of the number of string loop lensing
events expected in each survey, namely the angular resolution and
sensitivity.  The angular resolution sets the minimum separation of
images which may be detected, and hence the minimum string tension of
observable loops.  The sensitivity of the instrument influences the
number of potential targets for lensing, since a more sensitive
instrument will pick up fainter or more distant sources.

\subsection{Loops as lenses}\label{ss:LoopsAsLenses}
 
We rely on the approximation, justified below, that a loop of mass 
$M_{loop}$ produces multiple images of background sources situated in a 
patch of sky of area $\pi \theta_E^2$, and that the images are 
separated by at least $\theta_E$, where $\theta_E$ is the Einstein 
radius of a Schwarzschild lens having the same mass
as the loop.  The Einstein radius for a loop of mass $M_{loop}$ is given by
\be\label{eq:LoopSkyFractionApprox}
\theta_E = \sqrt{\frac{4GM_{loop}}{c^2} \frac{D_{LS}}{D_L D_S}}
\ee
where $D_{L}$, $D_S$ and $D_{LS}$ are the angular diameter distances 
between the observer and lens, observer and source, and lens and 
source, respectively \cite{Schneideretal1992}.

To justify this approximation we compare it to the 
sky area lensed by a circular loop of radius $R$ and angular radius 
$\theta_{loop} = R/D_L$ situated in a plane perpendicular to the line 
of sight.  Light rays passing outside the loop are deflected as if by a 
Schwarzschild lens of the same mass as the loop, while those passing 
through the loop are not deflected.  A pure Schwarzschild lens always 
produces two images of a source at angles $\theta_\pm$ from the
lens, given by 
\be\label{eq:SchwarzsLensImages} \theta_\pm = \frac{\theta_S \pm 
\sqrt{\theta_S + 4\theta_E^2}}{2},
\ee 
where 
$\theta_S$ is the (unlensed) angular position of the source.
For a loop, any images at $\theta < \theta_{loop}$ are spurious, since 
the corresponding light rays must have passed through the loop, and 
will not be observed.  The unlensed source itself will be visible 
through the loop if $\theta_S < \theta_{loop}$.

\begin{figure}[htb]
\begin{center}
\resizebox{\columnwidth}{!}{\includegraphics[angle=90]{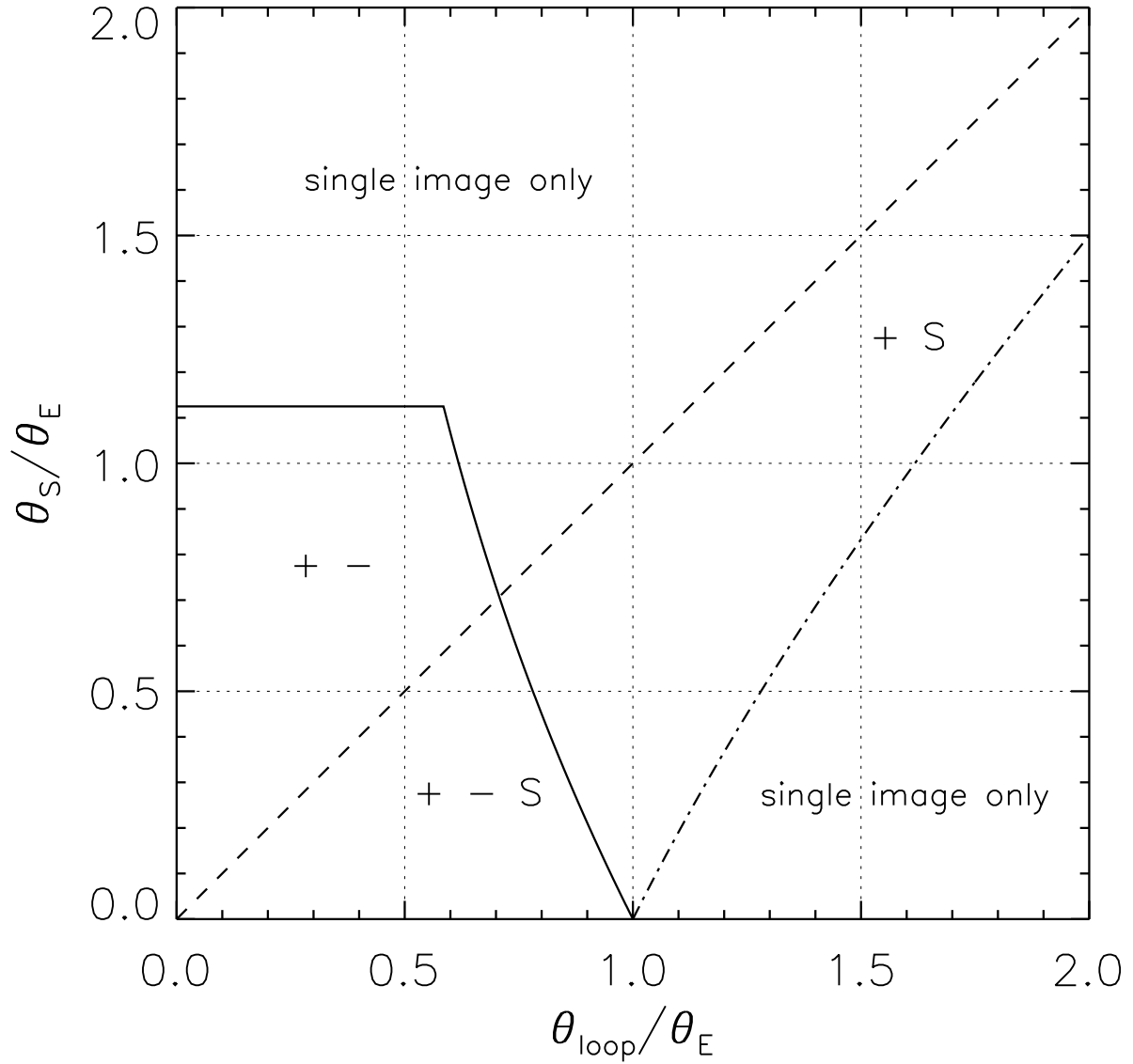}}
\end{center}
\caption{Regions of the $(\theta_S,\theta_{loop})$ parameter space 
where multiple images are formed by a planar circular loop.  Regions 
are labeled depending on whether the $\pm$ Schwarzschild images are 
formed, or whether direct images of the source (S), are visible.}
\label{f:LoopLensing}
\end{figure}

The region of $(\theta_S,\theta_{loop})$ parameter space in which 
multiple images form is shown in Figure \ref{f:LoopLensing}. While 
multiple images are always created by a Schwarzschild lens, the ``$+$'' 
and ``$-$'' images are produced with different fluxes, and if the flux 
ratio is too large then the pair will not be observed as a lens. 
Imposing a conservative flux ratio $ \leq 10$:1 for our calculations 
means that the ``$-$'' image 
will only be included if $\theta_S < 1.125 \,\theta_E$, which cuts off 
the multiple lensing region for the ``$-$'' image.  The source
and ``$+$'' image have a flux ratio that is nearly unity, which enhances 
the prospects for lens identification when both are present.  When 
$\theta_{loop} \leq \theta_E$ there is always at least one pair of 
images separated by $\ge 2\theta_E$.  For $\theta_{loop} > \theta_E$, 
the image separations gradually decrease from $\theta_E$ when 
$\theta_{loop} \sim \theta_E$, and go as $\theta_E^2/\theta_{loop}$ for 
$\theta_{loop} \gg \theta_E$.  The patch of sky for which multiple 
images form when $\theta_{loop} > \theta_E$ has area
\be\label{eq:BigLoopsSkyArea}
\pi\theta_E^2\left(2-[\theta_E/\theta_{loop}]^2\right)
\ee
which is always greater than $\pi\theta_E^2$.
One can show that the sky area (\ref{eq:BigLoopsSkyArea})
asymptotes to that of a long straight string as $\theta_E/\theta_{loop} 
\to 0$, corresponding to $R\to\infty$.

Our approximation accurately reflects the lensing cross section in the
special case of a planar circular loop.  However, the area lensed by a
typical loop most likely exceeds this.  A circular loop
is the least dense arrangement of the 
loop's mass, and furthermore rays passing through the loop are not 
lensed at all.  A realistic loop will likely be ``crumpled,''
concentrating the mass and better approximating a Schwarzschild lens
for a larger range of impact parameters.
In the limit that the loop can be treated as a random walk with step 
size (correlation length) $l_c$, then a loop of radius $R$, when 
crumpled, will fit in a sphere of radius $C\sqrt{l_c R}$, with $C$ a 
constant of order unity. 
This reduces the angle $\theta_{loop}$ 
subtended by the loop by the factor $C\sqrt{l_c/R}$, which will be 
significant for large loops.  There are other effects that may also 
increase the area of the lensed patch of sky, which we will discuss in
Section \ref{s:Discussion}.

\subsection{Loop population models}\label{ss:LoopPopModels}

We now describe our two models of the loop length distribution, defined 
by a number density per logarithmic interval in the loop length $L$, 
$N_L(L,z) = \ud N(z)/ \ud \ln L$.  Here $N$ is the number of loops in
a comoving 
volume $(c/H_0)^3$, where $c$ is the speed of light and $H_{0}$ is
today's Hubble parameter. 
The first model is the simplest scale-free 
distribution, a powerlaw in the loop length.  It is not motivated by any
model of string network evolution that we are aware of, but is useful for
understanding how the constraint depends on basic features of the
loop length spectrum and may describe other lens populations.
The second model is more realistic.  It uses the loop length spectrum
derived from the one-scale model (OSM) of cosmic string network evolution, 
a semi-analytic model that fits well to full-scale simulations.

\subsubsection{Powerlaw loop spectrum} \label{sec:powerlaw}

For the powerlaw spectrum, assuming a constant comoving density of 
loops, we have
\be
N_L(L,z) = \Omega_{loop,0} \frac{\rho_{crit,0} |\gamma|}{\mu L_*} 
\left(\frac{c}{H_0}\right)^3
\left(
\frac{L}{L_*} \right)^{\gamma-1},
\label{eqn:powerlaw}
\ee
where $\Omega_{loop,0}$ is the present mass parameter in loops, and 
$\rho_{crit,0}$ is today's critical density.  $L_*$ is a loop length 
cutoff, so we include only $L<L_*$ when $\gamma>0$, or $L>L_*$ when 
$\gamma<0$. If we instead assume a constant number of loops per 
particle horizon volume we obtain
\be 
N_L(L,z) = \Omega_{loop} \frac{\rho_{crit,0} |\gamma|}{\mu L_*} 
\left(\frac{c}{H_0}\right)^3
\left(
\frac{L}{L_*} \right)^{\gamma-1} \left[ \frac{H(z)}{H_0} \right]^3.
\label{eqn:case2}
\ee
To better compare with the one-scale model (discussed in the following
section) we will use this assumption (equation \ref{eqn:case2}) in all
our calculations with the powerlaw model henceforth.

For the present work we will only consider $\gamma<0$, as suggested by 
string network simulations \cite{Vanchurinetal2006,Ringevaletal2005}.
The effective lower cut-off 
on the powerlaw loop spectrum then comes from the smaller of $L_*$ and 
the resolution limit.  In practice, $\Omega_{loop}$ and $L_*$ are 
degenerate in the sky fraction calculation, so for this work we fix 
$L_* = 1$ kpc and vary $\Omega_{loop}$ only.  The upper limit on the 
loop spectrum integration is chosen so that loops longer than $L_{max} 
= (\epsilon/(1+\epsilon))^{1/\gamma} L_*$, which account 
for a fraction $\epsilon$ of the total number of loops, are left
out. In cases where 
the instrumental resolution is high enough to resolve loops all the way 
down to the lower cut-off $L_*$, defining the upper cut-off in this way 
results in the sky fraction being independent of $\gamma$.

\subsubsection{One-scale model} \label{sec:osm}

The one-scale model is a simple model of string networks, supported by 
numerical simulations: a detailed exposition may be found in 
\cite{CaldwellAllen1992}. The only length scale in this model is 
$\ell(t)$, the radius of the particle horizon at a time $t$.  Loops 
that form at time $t$ have length $\alpha \ell(t)$, with $\alpha$ a 
constant, and subsequently shrink by emitting gravitational radiation.  
The rate of loop formation is found by requiring that a sufficient 
number of loops are formed to keep the number of long strings per 
particle horizon volume constant. In principle, $\alpha$ can be 
determined from numerical simulations, but as yet a clear consensus on 
its value has not emerged, so we take it to be a free parameter.

By assuming a matter-dominated universe one finds
\be\label{eq:OSMLoopLengthSpectrum}
\frac{\ud N}{\ud \ln L} =
  \frac{1+\frac{\alpha/[G\mu/c^2]}{23.5}}{50.3\alpha [G\mu/c^2]}
  \frac{L/\ell_{H_0}}{\left( 1+\frac{L}{33.3[G\mu/c^2]\ell_{H_0}}
  \right)^2}
\ee
where $\ell_{H_0} = c/H_0$, and we have used the values for other 
one-scale model parameters in \cite{CaldwellAllen1992}. This 
distribution has a peak at $L_{peak} \sim 33 \ell_{H_0}[G\mu/c^2]$, and 
the spectral index $\gamma =0$ for $L > L_{peak}$ and $\gamma=+2$ for 
$L < L_{peak}$.  It has an upper cutoff at $L = \alpha \ell(t) =
  2 \alpha \ell_{H_0}$ (during the matter era); this is the length at
which loops are created.  Using this spectrum, 
the density parameter in loops is 
\be\label{eq:OSMOmegaLoop}
\Omega_{loop} = \left[ \frac{G\mu}{c^2} \right] \frac{185}{y} \left(
  1+\frac{y}{23.5} \right) 
  \left[ \ln \left( 1+\frac{y}{16.7} \right)
  - \left( 1+ \frac{16.7}{y} \right)^{-1} \right]
\ee
where  $y \equiv \alpha/(G\mu/c^2)$.  These equations are derived in detail in
Appendix \ref{s:OSMDerive}.

\subsection{Compact Radio Sources}\label{ss:CRS}

Selection of a background source population has a major impact on
the efficiency and completeness of a gravitational lensing survey.
For many background sources with extended emission, multiple imaging 
can be mistaken for the intrinsic structure of the source, or vice-versa.  
In order to easily determine whether or not lensing has occurred, the best 
strategy is to optimize the survey to observe sources that are intrinsically
point-like (to prevent the misidentification of substructure) and to
use instruments with resolution high enough to distinguish the images
produced by a typical lensing event.  With this in mind, we consider a
source population of 
flat spectrum compact radio sources (CRSs), which are
single-component sources at $\sim 100$ mas resolution.  These
sources are numerous at high redshifts; typically $z\sim1-2$ over
a few magnitudes in 
flux density (see, e.g.,
\cite{Marlowetal2000,Falcoetal1998}). This enables 
us to probe a large cosmological volume.
Compact radio sources are also ideal for the 
purpose of optimizing the survey resolution, as long-baseline radio 
interferometry routinely reaches sub-arcsecond resolution. 
Furthermore, radio sources may be significantly polarized, offering a 
further check that lensing is indeed occurring, as well as providing
additional constraints on lens models.

The signatures of lensing by cosmic string loops would be quite
different from 
those seen in lensing by galaxies. In every instance of lensing by a galaxy, 
the lens would be identified by its presence in follow-up optical or near-infrared 
observations (normal elliptical and spiral galaxies are fairly radio quiet). In
addition, lens modeling would indicate whether the lens has a 
mass density profile compatible with a galaxy, the majority of which have near isothermal
profiles on the scales relevant to strong lensing \cite{DobkeKing06}.
The polarization of the images would also help to differentiate between lensing by
galaxies and by cosmic string loops.
In galaxy lensing, any
polarized emission of the source is affected by Faraday rotation as it passes through the
interstellar medium of the lens galaxy, resulting in (frequency-dependent)
differences in the polarization position angles of the images; this would not occur if
the lens was a cosmic string loop.
Using these criteria it will be straightforward to rule out lensing by
a galaxy, and careful follow-up 
imaging and spectroscopic observations will make it possible to determine if the lens is any
known astrophysical object.

\subsection{Radio Surveys}\label{ss:RadioSurveys}

Having described the fraction of sky that could be lensed by string 
loops, and having selected our source population, we now discuss 
some instruments and surveys that might observe loop lensing events.  
Key features that will serve as inputs to the estimated number of 
lensing events are the dynamic range, resolution, and expected source 
counts for each survey.

\noindent
{\bf CLASS/JVAS.} The Cosmic Lens All-Sky
Survey (CLASS) \cite{Browneetal2003,Myersetal2003}
and the Jodrell Bank VLA Astrometric Survey (JVAS) \cite{Kingetal1999}
searched for gravitationally lensed flat spectrum radio sources. Together,
they targeted 16503 sources, of which 11685 have flat spectra (spectral
index flatter than 0.5 between 1.4 and 5 GHz) and flux density $\ge$ 30mJy
at 5 GHz.
The statistically well defined sample of 11 lens systems found
includes only those with separations in excess of 0.3-arcsec and flux 
ratio $\le$ 10:1.

\noindent
{\bf LOFAR.} The Low Frequency Radio Array (LOFAR) is expected to be 
operational by the end of 2008, initially with a maximum baseline of 
100 km, and maximum resolution of $\sim$3 arcsec, which would only 
detect the largest separation events. Therefore, we concentrate on a
proposed extension to LOFAR with a maximum baseline of 400 km
\cite{Vogt2006}.
Two planned surveys are relevant to lensing: one at 200
MHz over 250 deg$^2$ at 0.8-arcsec resolution, and the other at 120
MHz over half the sky with a resolution of 1.3
arcsec \cite{Wucknitzetal2006}. 
The former will include 3$\times 10^{7}$ 
sources to the 14$\mu$Jy limit and the latter 8.6$\times 10^{8}$ 
sources to 43$\mu$Jy. These source counts are expected to be dominated 
by steep spectrum galaxies, with $\sim 5\%$ being flat-spectrum
\cite{Cohen2005}.

\noindent
{\bf SKA.} The Square Kilometre Array (SKA),
planned to be fully operational by 2020,
will consist of thousands of radio antennae with a collecting area
exceeding thirty times that of the largest telescope ever constructed. The
site and specifications have yet to be finalized; we adopt the
estimates from \cite{Koopmansetal2004}. They
discuss a possible Radio All-SKA Lens survey (RASKAL) of
half the sky at 1.4 GHz, with 0.01-0.02-arcsec resolution. Limiting
to sources brighter than 3$\mu$Jy there would
be $\sim 10^9$ radio sources, of which roughly 10$\%$ would
be compact, flat-spectrum sources -- the ideal target for a statistically
complete lens survey. As with CLASS/JVAS, the median redshift is expected
to be $z>1$.

We summarize the properties of the surveys we consider in Table
\ref{tab:survey}.  For Extended LOFAR, we take the higher-resolution
240 MHz survey, and for SKA we assume the optimistic end of the
resolution range, 0.01 arcsec.  In the table, ``number of sources''
refers to the number of CRSs accessible to the survey.
Based on estimates in \cite{Marlowetal2000} of the mean redshift of CRSs,
for the purpose of this study we choose a redshift of $z=1.2$ for our
source population in all surveys.  The true distribution of CRSs is
still incompletely understood, especially for the flux limits that
can be reached by future surveys.  We expect that this approximation
affects our result by less than an order of magnitude.

\begin{table}[h!b!p!]
\caption{Properties of Gravitational Lensing Surveys}
\begin{tabular}{|l|r|r|} 
\hline
Survey&Resolution&Number of Sources\\
&(arcsec)&(observed/expected)\\
\hline
CLASS/JVAS&0.3&11685\\
Ext. LOFAR&0.8&$1.5 \times 10^6$\\
SKA&0.01&$10^8$\\
\hline
\end{tabular}
\label{tab:survey}
\end{table}

\section{Results}\label{s:Results}

\subsection{Expected Lensing Events}

We estimate the number of cosmic string loop lensing events one can
expect to observe in past and future compact radio source
gravitational lensing surveys whose properties can be found in
Table \ref{tab:survey}. Our main results are
summarized in Figures \ref{fig:powerlaw} and \ref{fig:osm}.

\subsubsection{Powerlaw models}
\begin{figure}[htb]
\begin{center}
\resizebox{\columnwidth}{!}{\includegraphics[angle=90]{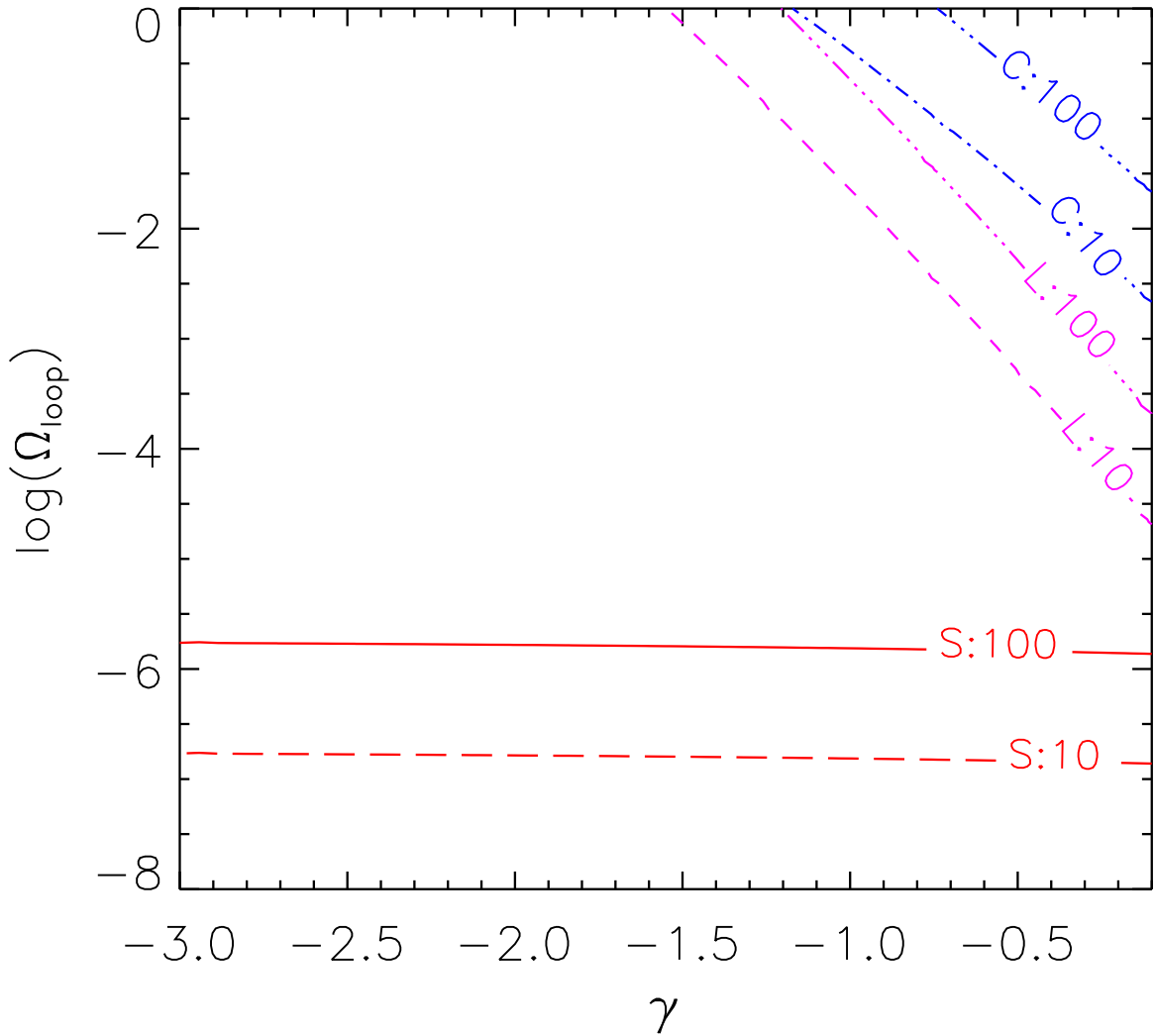}}
\end{center}
\caption{Contours for 10 and 100 expected events using the powerlaw
  loop spectrum model for CLASS/JVAS (C), LOFAR (L) and SKA (S), for a
  source population at redshift z=1.2.
  The x-axis is the powerlaw index $\gamma$ (see Section
  \ref{sec:powerlaw}) and the y-axis is the density parameter in
  loops, $\Omega_{loop}$. 10 expected events corresponds to a
  constraint at 3$\sigma$ if no events are observed.} 
\label{fig:powerlaw}
\end{figure}

For a powerlaw loop length spectrum with a negative spectral index,
most loops will be small.
This is a disadvantage in a lensing search because of image
separation: lensing events from small loops are likely to be lost
due to resolution limits.  With CLASS/JVAS, we expect to
see few lensing events in the survey over most of the range
of plausible powerlaw indices and loop density parameters.  LOFAR will
have the ability to rule out a slightly larger portion of the
parameter space,
for $\Omega_{loop} \gtrsim 10^{-2}$ and $\gamma \gtrsim -1$.  When
$\gamma<0$, the sky fraction is strongly 
dependent on the instrument's ability to resolve the smallest loops;
both CLASS/JVAS and LOFAR lack the resolution to be effective in this
regime.  With the high resolution of SKA, however, nearly all loops
will be resolved
for powerlaw indices $-3 < \gamma < 0$ with $L_* = 1$ kpc, and the
large number of compact radio sources available will allow us 
to rule out to 3-$\sigma$ a density parameter in loops of $\sim
10^{-7}$ if no loop lensing events are observed.  The results for the
powerlaw loop spectrum are presented in Figure \ref{fig:powerlaw}.
(Note that these results could easily be translated into constraints
on any population of point masses with a powerlaw mass spectrum, which
may include some models of primordial black holes~\cite{Carr1975}.)

\subsubsection{One-scale model.}
\begin{figure}[htb]
\begin{center}
\resizebox{\columnwidth}{!}{\includegraphics[angle=90]{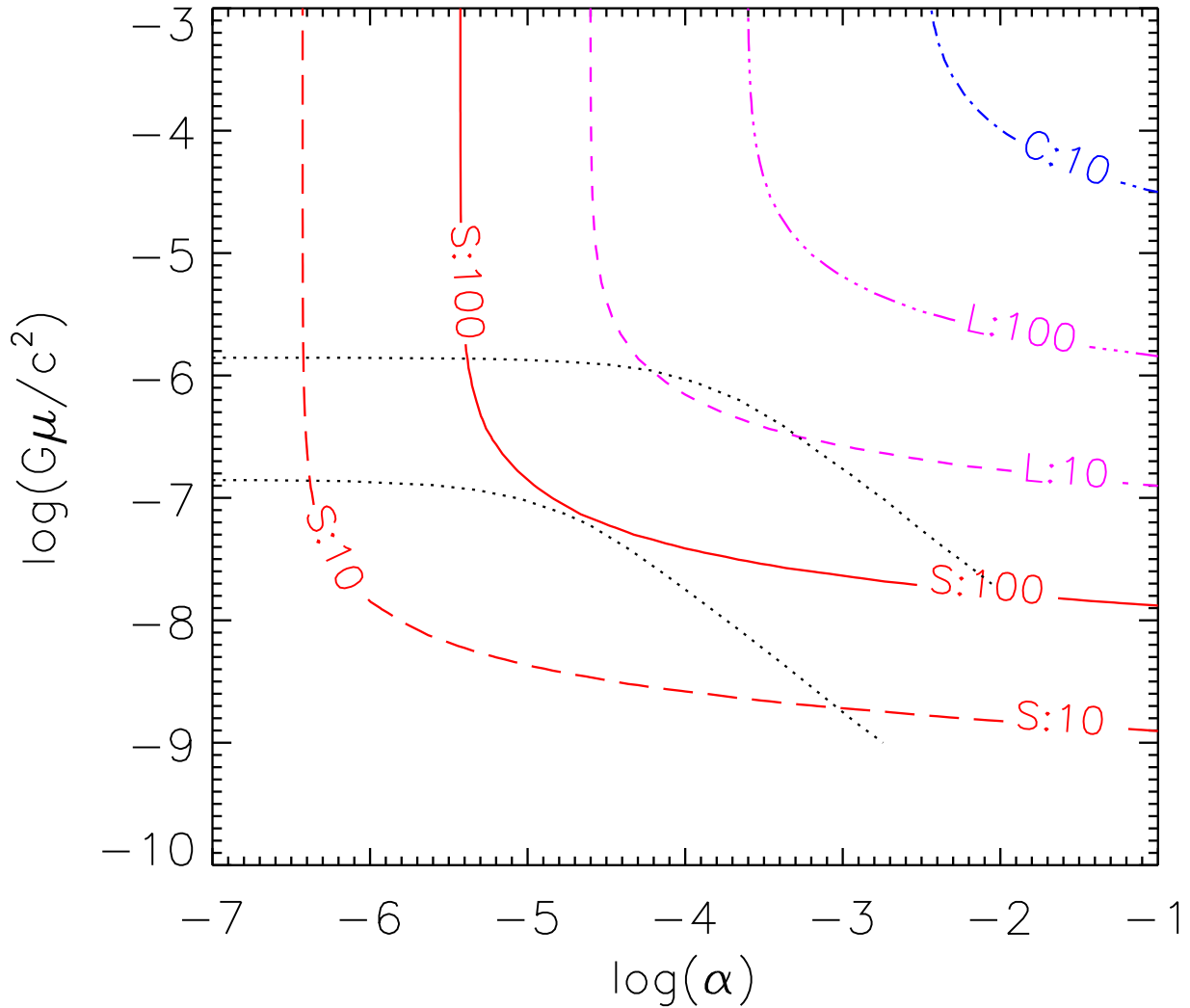}}
\end{center}
\caption{Contours for 10 and 100 expected events using the one-scale
  loop spectrum model for CLASS/JVAS (C), LOFAR (L), and SKA (S), for a
  source population at redshift z=1.2. We also plot the pulsar timing
  constraint from \cite{Battyeetal2006} (black dotted lines).  The upper
  dotted line is for $\Omega_g h^2 < 2 \times 10^8$ and the lower is for
  $\Omega_g h^2 < 2 \times 10^9$ -- see \cite{Battyeetal2006} for a
  discussion of the uncertainty.  In this plot, the
  x-axis is the one-scale 
  model parameter $\alpha$ (see Section \ref{sec:osm}) and the y-axis
  is the dimensionless loop tension $G\mu / c^2$.}
\label{fig:osm}
\end{figure}

Our results for a one-scale model spectrum of
loops are presented in Figure \ref{fig:osm} for the CLASS/JVAS, 
LOFAR and SKA surveys.
The one-scale model spectrum has two limiting cases of interest,
depending on the relationship between $G\mu / c^2$ and $\alpha$.

When $\alpha \lesssim 16 G\mu/c^2$, the upper cut-off in the spectrum is
below the peak, and thus the spectrum is approximately a powerlaw with
spectral index +2 and is dominated by loops near the upper cut-off at $2
\alpha \ell_{H_0}$.  For a powerlaw spectrum as in Equation
\ref{eqn:powerlaw} with $\gamma > 0$, the mass density goes as
\be
\rho \sim \left( \frac{G\mu}{c^2} \right) N_0
L_{max}^{1-\gamma}(L_{max}^\gamma-L_{min}^{\gamma}),
\ee
with $N_0 \sim \Omega_{loop}/([G\mu/c^2]L_{max})$.  This means that
the mass density $\rho \sim \Omega_{loop} \sim 1/\alpha$.
Ignoring resolution
effects, the sky fraction is proportional to the mass density, since
$\theta_E^2 \sim M_{loop}$, so in this regime, the sky fraction is
independent of $G\mu/c^2$.  When $\alpha$ is small, the resolution of
the instrument also decreases the sky fraction because the smallest
image separations will be lost.

When $\alpha >> G\mu/c^2$, the mass in loops is dominated by the
region of the spectrum with a spectral index of 0.  For $\gamma = 0$,
$\rho \sim (G\mu/c^2)N_0 \ln{(L_{max}/L_{min})}$.  Since the low-mass end
of the spectrum gives a small contribution, one can approximate the
lower cut-off as the peak position: $L_{min} \sim L_{peak} \sim
G\mu/c^2$.  $L_{max}$ is the upper cut-off which goes as $\alpha$.  In
this case, $N_0 \sim \Omega_{loop}/([G\mu/c^2]
\ln{\alpha/[G\mu/c^2]})$.  Since in the limit that $\alpha >>
G\mu/c^2$, $\Omega_{loop} \sim (G\mu/c^2) \ln{\alpha/(G\mu/c^2)}$, one
finds that $\rho \sim (G\mu/c^2) \ln{\alpha/(G\mu/c^2)}$.  In this
regime, the sky fraction depends only very weakly on $\alpha$.

If the loop spectrum follows the one-scale model, we can rule out
regions of the parameter space to 3-$\sigma$ even with the CLASS/JVAS
survey, as seen in Figure \ref{fig:osm}.  This plot also
shows that a large region of parameter space
can be ruled out with LOFAR or SKA.  With SKA, a 3-$\sigma$ constraint
will be achievable down to $G\mu/c^2 \sim 10^{-9}$ for large values of
$\alpha$.  In the next section, we discuss
how loop lensing surveys can be more powerful and more direct than
current methods to constrain cosmic strings with pulsar timing.

\subsection{Comparison with other constraints}

Currently, the strongest limits on cosmic strings come from constraints 
on the stochastic gravitational wave background obtained from the 
timing of millisecond pulsars \cite{DePiesHogan2007}.  In the context
of the one-scale model, pulsar timing constrains a combination of
$G\mu/c^2$ and $\alpha$.

In Figure \ref{fig:osm}, we include the pulsar timing constraint from
\cite{Battyeetal2006} for two different values of the density parameter
in gravitational waves for comparison with the limits we can obtain with
radio surveys.
The limit obtainable from SKA can improve 
upon the pulsar timing constraint by up to two orders of magnitude in 
$G\mu/c^2$ in some regions of the parameter space.

However, a simple comparison between limits from the gravitational wave 
background and from loop lensing surveys can obscure the most important 
feature of a lensing survey, which is that it is a {\it direct search}.
While constraints from pulsar timing rely on models of the loop
radiation and simulations of the behavior of loop networks, lensing
relies only on the presence of a mass concentration, the effects of
which are directly observed in the image distributions.  Any model of
a cosmic string population will make simple and testable predictions
for gravitational lensing surveys.

\section{Discussion}\label{s:Discussion}

We have found that searching for lensed compact radio sources in future 
surveys can greatly improve constraints on cosmic string networks.
Extending the excluded region of parameter space, or alternatively
observing a cosmic  string, will allow us to better understand physics
beyond the Standard Model.
We have shown that if no cosmic string loop lensing events are
observed in upcoming radio surveys, much of the currently interesting
parameter space can be ruled out, but
further work is needed to determine what one could learn about the
nature of cosmic strings if a lensing event were actually to be observed.  For the sake of obtaining a constraint, it is sufficient to discuss the scenario in which all lensing events in a survey can be accounted for by the detection of a conventional lensing object (such as a galaxy or cluster) in follow-up observations, so there is no need to invoke cosmic strings.  In all known strong gravitational lensing events to date, observations have identified the source of the lensing potential as a standard baryonic or dark-matter structure.

In this work, we have tried to make conservative assumptions in order
to estimate the lensing signal, but including additional effects is
only likely to improve our limits.
For instance, the statistically complete sample of lenses in CLASS
\cite{Browneetal2003} includes only
those systems with flux ratio less than 10:1, but
future radio interferometers with higher dynamic range 
may be expected to easily improve on this ratio by at least an order of 
magnitude. This would significantly increase the effective lensing area 
for each loop to as much as an order of magnitude above the area within 
the Einstein radius.

In any survey, brighter events are more likely to be above the flux 
limit, and so lensing surveys will see a disproportionate number of 
high-magnification events.  By not explicitly including the effect of
this magnification bias in our estimate, we are
underestimating the number of lensed sources.

The detailed properties of the cosmic strings under consideration will
also likely enhance their prospects for detection.  One such property
is the reconnection probability $P$, which is unity for ordinary
cosmic strings \cite{Shellard1987}, but 
can be significantly less than unity for cosmic strings originating
from superstring models
\cite{JonesStoicaTye2003,Jacksonetal2004}.  Simulations indicate
that for $P < 0.1$, the string number density increases as $\sim
P^{-0.6}$ \cite{AvgoustidisShellard2006}, which would proportionally
increase the number of expected
lensing events.  In addition, there may be several
distinct string populations present: for example, in brane inflation
scenarios multiple types of cosmic strings, labeled by coprime
integers $(p,q)$, would be produced.  The total number of strings
would be enhanced by the number of different populations present
\cite{Tyeetal2005}, enhancing the lensing signal by the same
factor. 

Further work on the statistical properties of loop lensing will help 
to sharpen our constraint.  We expect that considering lensing by
loops in more general configurations (including oscillations) will
only slightly affect the total lensing cross section.  However,
understanding the effects of configuration may be important for
determining the detailed properties of individual loops detected in
lensing surveys. 

Experiments such as LISA will revolutionize the observation of
gravitational waves, in particular the stochastic background, to which
cosmic strings are expected to contribute a distinctive signature
\cite{Hogan2006}.
In comparison with limits from millisecond pulsars, LISA
is expected to push back the minimum detectable string threshold by
$\sim 7$ orders of magnitude \cite{DePiesHogan2007}. Although
these tensions are inaccessible to the surveys we mention in this work,
a lensing search is a complementary technique in that in addition to
requiring fewer assumptions about the behavior of strings, it involves
a local phenomenon rather than a spectral signature from a
population.

We hope that our results will motivate the optimization of future
surveys for events of this nature.
The next generation of radio surveys have the potential to directly
detect cosmic string loops, opening a new window into the physics of
the early universe. 

\section{Acknowledgements}

This work was supported in part by a NSF Graduate Research Fellowship
(KJM).  KJM would like to acknowledge useful discussions with 
George Efstathiou, Rob Fender, Tom Maccarone, David Spergel and
Paul Steinhardt and the
hospitality of the IoA and DAMTP (Cambridge University) during the
time this work was carried out.  LJK is supported by the Royal
Society.  She also thanks Antony Lewis, Guy Pooley and Elizabeth
Waldram for useful
discussions.  DHW thanks Mark Wyman for informative discussions, and
the Perimeter Institute for its hospitality during the final stages of
this work. 

\appendix

\section{One-scale model loop spectrum}\label{s:OSMDerive}

In this Appendix we describe the process by which the loop length distribution
(\ref{eq:OSMLoopLengthSpectrum}) and total mass in loops (\ref{eq:OSMOmegaLoop}) are 
derived in the context of the one-scale model (OSM).  
The equations and parameter values we use are those of \cite{CaldwellAllen1992}.
According to the
OSM, the energy density in long strings scales like $\ell^{-2}$ where
$\ell$ is the particle horizon, defined by
\be
\ell(t) = a(t)\cdot \int_0^t \frac{\text{d}t'}{a(t')} = 3ct,
\ee
and the last equality holds during matter domination.
Applying energy conservation then determines the rate
at which the long string network sheds energy in the form of loops. The
loops formed at time $t$ are all taken to have an initial length $\alpha \ell(t)$, where $\alpha$
is a dimensionless constant.  These assumptions imply that the 
absolute number $N$ of
loops created in a comoving cubical volume of size $R$ on a side is given by
\be\label{eq:OSMLoopCreationRate}
\frac{\ud N}{\ud \ell} = \frac{a(t)^3 R^3}{\ell^4} \cdot \frac{C}{\alpha}
= \frac{C}{4\alpha} \frac{\ell_{H0}}{\ell^2}
\ee
where $C = 5.3$ during matter domination.  To obtain the
second equality, we have  taken $a=1$ at a fiducial time $t_0$, 
and used $a(t) = (t/t_0)^{2/3} = (\ell/2\ell_{H0})^{2/3}$, where
$\ell_{H0} = c/H(t)$ is the Hubble length at $t_0$.  We have further taken
 $R = \ell_{H0}$, so $N$ is the number of loops in a comoving (cubical) horizon volume.

Once loops have formed, they begin to shrink as they lose energy by emitting gravitational
radiation.  The energy loss rate is 
\be
\frac{\ud E}{\ud t} = - \Gamma G\mu^2c ,
\ee
where $\Gamma \sim 50$ is taken from simulations.  This means that the length of a loop
that was ``born" at time $t_B$ is
\be
L(t,t_B) = f_r \alpha \ell(t_B) - \frac{\Gamma G\mu}{c}(t-t_B),
\ee
where $f_r \sim 0.7$ is a factor accounting for the shrinking of the loop as its initial 
relativistic velocity is redshifted by Hubble expansion. During matter 
domination the time $t_D$ at which a loop ``dies" (completely evaporates) is related to its
time of birth by
\be
t_D = \frac{t_B}{\beta}, \quad
\beta = \left( 1 + \frac{3f_r\alpha c^2}{\Gamma G\mu} \right)^{-1}.
\ee
This also means that a loop of length $L$, observed 
when the particle horizon is $\ell$, was
born when the particle horizon was a size $\ell_B$ given by
\be\label{eq:OSMdelldL}
\ell_B = \beta \left( \ell + \frac{3 c^2}{\Gamma G \mu} L \right).
\ee
In our universe, if $\beta > 5 \times 10^{-6}$, then all the loops created during radiation domination will have evaporated by the present day.  We will assume this is
so in the following.

Our first goal is to calculate the spectrum of loop lengths, evaluated at the fiducial time $t_0$.  We proceed by writing
\be
L \frac{\ud N}{\ud L}
\bigg|_{t_0} = L \frac{\ud N}{\ud \ell_B} \frac{\ud \ell_B}{\ud L}
\ee
where $\ell_B$ is the particle horizon length when a loop currently of length $L$ was
created.  The first derivative on the right hand side is given by 
(\ref{eq:OSMLoopCreationRate}), and the second is obtained from
(\ref{eq:OSMdelldL}).  Combining them, we find
\be
L \frac{\ud N}{\ud L}
\bigg|_{t_0}
= 
\frac{3}{16}\frac{C c^2}{\alpha\beta\Gamma G\mu}
\frac{L/\ell_{H0}}{\left( 1 + \frac{L}{2\Gamma G\mu \ell_{H0}/3c^2}\right)^2},
\ee
which, once the values for $f_r,\Gamma$ and $C$ are substituted, becomes precisely
equation (\ref{eq:OSMLoopLengthSpectrum}).

To compute the density parameter in loops, we first calculate the total length
in loops.  This requires the integral identity
\be
\int_0^{L_*} \frac{AL/B}{(1+L/B)^2} \; \ud L
= AB \left[ \ln\left(1 + \frac{L_*}{B}\right) - \frac{L_*/B}{1+L_*/B} \right],
\ee
which applies to our case with the choice,
\be
A = \frac{C}{8\alpha\beta}, \quad
B = \frac{2\Gamma G\mu \ell_{H0}}{3c^2}, \quad
L_* = \alpha \ell = 2\alpha \ell_{H0}.
\ee
The upper cutoff is necessary because the largest possible loops are those being created
right at $t_0$.
The integral formula gives
\be
L^{(tot)}_{loop} = 
\ell_{H0} \cdot \frac{\Gamma C}{12 y} \left( 1 + \frac{3f_r}{\Gamma} y \right)
\left[ \ln \left( 1 + \frac{3}{\Gamma} y \right) - \left( 1 + \frac{\Gamma}{3y} \right)^{-1}
\right]
\ee
where we have taken $y=\alpha/(G\mu/c^2)$.
From the standard definition of the density parameter, we have
\be
\Omega_{loop} = \frac{8\pi}{3} \frac{L^{(tot)}_{loop}}{\ell_{H0}}
\left(\frac{G\mu}{c^2}\right).
\ee
Finally,  substituting the values of $f_r,\Gamma$ and $C$ we obtain equation (\ref{eq:OSMOmegaLoop}).

\end{document}